# The Impact of Artificial Intelligence on Strategic Technology Management: A Mixed-Methods Analysis of Resources, Capabilities, and Human-AI Collaboration

**Massimo Fascinari**

Università Telematica Internazionale Uninettuno, Rome, Italy

Email: m.fascinari@accenture.com

**Vincent English**

Longford International College, Università Telematica Internazionale Uninettuno, Rome, Italy

Email: vincent.english@longfordcollege.com

## Abstract

This paper investigates how artificial intelligence (AI) can be effectively integrated into Strategic Technology Management (STM) practices to enhance the strategic alignment and effectiveness of technology investments. Through a mixed-methods approach combining quantitative survey data (n=230) and qualitative expert interviews (n=14), this study addresses three critical research questions: what success factors AI innovates for STM roadmap formulation under uncertainty; what resources and capabilities organizations require for AI-enhanced STM; and how human-AI interaction should be designed for complex STM tasks. The findings reveal that AI fundamentally transforms STM through data-driven strategic alignment and continuous adaptation, while success depends on cultivating proprietary data ecosystems, specialized human talent, and robust governance capabilities. The study introduces the AI-based Strategic Technology Management (AIbSTM) conceptual framework, which synthesizes technical capabilities with human and organizational dimensions across three layers: strategic alignment, resource-based view, and human-AI interaction. Contrary to visions of autonomous AI leadership, the research demonstrates that the most viable trajectory is human-centric augmentation, where AI serves as a collaborative partner rather than a replacement for human judgment. This work contributes to theory by extending the Resource-Based View to AI contexts and addressing cognitive and socio-technical chasms in AI adoption, while offering practitioners a prescriptive framework for navigating AI integration in strategic technology management.

# 1. Introduction

The persistent challenge of underperforming technology investments poses a substantial problem for contemporary organizations. Research by Gartner reveals that less than half of organizations achieve expected value from their Information Technology investments, (Stegman, Guevara et al. 2023) an issue amplified by the increasing budgets allocated to technological initiatives. This failure rate in technology investments stems primarily from suboptimal Strategic Technology Management (STM) practices, particularly the misalignment between business objectives and technology strategies.

Strategic Technology Management, defined as the process of aligning an organization's technology with its corporate-level business strategies to create and sustain competitive advantage (Sahlman and Haapasalo 2012), has become increasingly complex in the digital age. The rapid integration of artificial intelligence across various domains presents both an unprecedented opportunity and a fundamental challenge for organizations seeking to enhance their STM structures and practices. As AI technologies mature from experimental applications to core business capabilities, organizations face a transformation for which many are inadequately prepared.

This research addresses a critical gap in literature: despite extensive work on AI's technical aspects and broader strategic applications, there exists no unified, integrated framework for its systematic embedment within STM practices. This deficit highlights a fundamental challenge - how to transition from conceptual understanding of AI's capabilities to pragmatic and strategic implementation within existing operational frameworks. The significance of this research stems from recognizing AI not merely as another technological tool, but as a catalyst requiring a fundamental shift in how organizations approach strategic technology management.

The research is motivated by a central paradox: while AI promises to address human attention scarcity and enhance decision-making capabilities, it simultaneously creates new demands for human oversight, particularly in verifying outputs for complex strategic tasks. This tension is further complicated by divergent perspectives on AI's ultimate role. One perspective positions AI as augmenting human capabilities, enabling strategists to focus on higher-value activities like ethical oversight and creative problem-solving. Conversely, a more transformative vision explores pathways toward greater automation through AI agents, exemplified by concepts like the "Agentic CTO.

The primary aim of this study is to identify how organizations can leverage AI to enhance the value of their technology investments by analysing what barriers AI can reduce to improve managerial productivity and their ability to implement successful strategic technology management roadmaps. This investigation addresses three interconnected research questions:

**RQ1:** What critical success factors does AI innovate for formulating the STM roadmap under conditions of creativity and uncertainty?

**RQ2:** What resources and capabilities should organizations have to enhance their STM roadmap formulation in complex scenarios with AI?

**RQ3:** How should the interaction between humans and AI be designed to enhance the performance of complex STM tasks?

Through a comprehensive mixed-methods approach combining literature review, quantitative survey data, and qualitative expert interviews, this research develops the AI-based Strategic Technology Management (AIbSTM) conceptual framework. This framework serves as both a theoretical contribution and practical guide, synthesizing the technical capabilities of AI with the essential human and organizational aspects of STM.

The paper is structured as follows: Section 2 reviews relevant literature across three core themes - STM success factors, the Resource-Based View applied to AI, and human-machine interaction dynamics. Section 3 outlines the mixed-methods methodology. Section 4 presents findings from both quantitative and qualitative phases. Section 5 discusses these findings in relation to the research questions. Finally, Section 6 concludes with theoretical and practical implications for the field of strategic technology management in the AI era.

## 2. Literature Review

### 2.1 Strategic Technology Management and Success Factors

Strategic Technology Management involves planning, organizing, leading, and controlling technology implementations to support organizational business strategy (Roberts 2001). It contributes to formulating and executing long-term goals by allocating necessary strategic technology resources (Sahlman 2010, Deutsch and Berényi 2023). STM differs fundamentally from both technology strategies, which focuses on adopting specific technologies at the business unit level, and technology management, which concerns developing technological capabilities for operational goals (National Research Council 1987).

The literature identifies several critical success factors for effective STM. Strategic technology decisions prove most effective when aligned with business and competitive strategies, ensuring technology choices provide significant cost advantages or create highly valued stakeholder capabilities (Sahlman and Haapasalo 2012). Success is influenced by broader organizational factors including industry context, inter-organizational strategies, cross-functional collaboration, and human resource management practices (Meer and Calori 1989). Additionally, competitive environment, management control style, and incentive systems play significant roles in guiding strategic choices (Sands 1991).

Effective leadership and supportive organizational culture emerge as integral success factors. Leaders guide problem definition and solution development while establishing change-receptive cultures. However, research often overlooks how leader behaviours, collaboration, and processes directly influence organizational technology innovation (Kurzhals, Graf-Vlachy et al. 2020). Cooper (2024) advocates for a "four-point dialectic" that evenly distributes attention among strategist, technology, customer, and industry trends. Successful technology adoption requires integrating new technologies with existing culture (Borkovich, Skovira et al. 2015) and accumulating knowledge through experience and "learning by doing" to enhance technological capabilities over time (Kharbanda 2001).

The proliferation of large language models and evolution of AI-based agents increasingly support strategic alignment, particularly in volatile, uncertain, complex, and ambiguous (VUCA) environments (Krishnan 2025). These systems, capable of simulating scenarios and analysing vast datasets, redefine how organizations achieve strategic alignment (Holmström and Carroll 2024, Alexander D'Amico 2025). AI transforms alignment by linking organizational goals and technology roadmaps through data, enhancing adaptability through scenario simulation. Its predictive capabilities improve decision-making, enabling organizational ambidexterity - exploring new opportunities while exploiting existing strengths (Daskalopoulos and Machek 2025).

A core AI benefit is accelerating insight generation from extensive datasets, enabling bold, evidence-based strategies moving beyond intuition (Biloslavo, Edgar et al. 2024). Real-time data analysis and predictive foresight maintain alignment in rapidly changing environments (Schrage, Kiron et al. 2024). AI also addresses operational challenges like "blank page syndrome" by generating creative starting points and structuring unstructured data, preserving human judgment while enhancing dynamic decision-making capabilities (Yun, Feng et al. 2025).

## 2.2 Resource-Based View and AI Capabilities

The Resource-Based View (RBV) posits that firms achieve sustained competitive advantage through possessing and deploying resources that are valuable, rare, inimitable, and non-substitutable (VRIN) (Wernerfelt 1984, Barney 2001). This framework proves particularly relevant for analysing AI adoption, as its uneven implementation across firms can be explained by differences in access to AI-related VRIN resources and capabilities.

Resources encompass tangible or intangible assets firms own or control, while capabilities represent the ability to effectively combine, deploy, and leverage these resources strategically. AI requires specific resources including powerful computational hardware, natural language processing capabilities (Krishnan, Elmore et al. 2019), and large datasets for effective decision-making (Jones and Wray 2006, Massoudi, Fatah et al.

2024). However, resources alone prove insufficient. Organizations must develop capabilities enabling collaboration with AI researchers through open innovation while retaining distinct internal execution strategies (Haefner, Wincent et al. 2021, Agarwal 2025).

The scaling law of AI performance critically influences resource requirements, positing that improvements depend on jointly expanding data, compute, and model size (Kaplan, McCandlish et al. 2020). This creates resource asymmetries predicted by RBV to underpin competitive advantage. However, when AI capabilities are commoditized as services, smaller firms can compete by leveraging unique data niches, domain expertise, or inimitable governance routines (Alexander D'Amico 2025, Diyin and Bhaumik 2025).

Governance capabilities emerge as particularly critical, encompassing data-driven workflow management, role definition, human-AI collaboration promotion, and strategic AI positioning (Perifanis and Kitsios 2023). Effective governance necessitates data ownership, curation capabilities, ethical frameworks, risk assessments, and lifecycle compliance automation (Floridi, Buttaboni et al. 2025, IBM 2025). The EU AI Act exemplifies how governance may transition from differentiating capability to market prerequisite.

Human talent represents another differentiating resource, valuable for specialized, often tacit knowledge. Integrating AI with strategic management demands "fusion skills" combining domain expertise with AI literacy (Purdy and Williams 2023, Mäkelä and Stephany 2024). Middle management must acquire AI skills to bridge communication between AI adopters and inhibitors (Rowe, Suire et al. 2024). Organizations require multi-pronged approaches including targeted hiring, reskilling, and workforce reconfiguration to position AI as an effective differentiator.

Organizational culture proves foundational for success. A data-centric culture establishes robust governance for quality and privacy while promoting data-driven decision-making. Agile cultures respond effectively to AI-driven changes, developing digital competencies enabling effective AI tool usage (Čižo, Komarova et al. 2025, Li 2025). Cultural readiness requires proactive change management and AI literacy programs securing stakeholder buy-in (Abdullah , Kumar, Ishan et al. 2025).

Dynamic capabilities - organizational processes enabling environmental adaptation (Teece, Pisano et al. 1997) - prove essential for sustained AI advantage. These include sensing opportunities, seizing resources for promising initiatives, and transforming workflows and structures (Owusu and Agbesi 2025). Self-improvement capabilities represent an advanced frontier, with techniques like automated quality control loops and AI-enabled R&D allowing dynamic strategy adjustments (Lu, Lu et al. 2024, Kokotajlo, Alexander et al. 2025).

## 2.3 Human-AI Interaction in Strategic Contexts

Integrating AI into organizational decision-making introduces significant complexities requiring well-thought-out strategies (Davenport 2021, Kesting 2024). The boundary between AI capabilities forms a "jagged technological frontier" where optimal approaches remain incompletely understood (Dell'Acqua, McFowland et al. 2023). Technology leaders play crucial roles guiding organizations through shifts to cognitive economies where intelligent machines become workforce-central (Naqvi 2017).

AI impact depends fundamentally on human-AI relationships, requiring understanding of motivations, emotions, behaviours, and attitudes (Chernov, Chernova et al. 2020, Haefner, Wincent et al. 2021). Human-AI symbiosis models must include safeguards preventing task polarization, where simple tasks become simpler while complex tasks grow more complex (Simkute, Tankelevitch et al. 2024). This creates paradoxes where human attention verifies AI outputs for complex tasks despite AI addressing attention scarcity (Eriksson, Bigi et al. 2020, Woodruff, Shelby et al. 2024).

Effective integration necessitates significant organizational and cultural adjustments through leadership-driven approaches addressing operational transformation and interaction dynamics (Schrage, Kiron et al. 2024). Success requires developing KPIs assessing AI integration value, moving beyond ad hoc applications toward strategic frameworks. Culturally, organizations must address leadership resistance by demonstrating AI value through executive training enabling collaborative data interrogation and scenario exploration.

According to Purdy and Williams (2023), effective human-AI collaboration relies on four considerations: domain-specific AI applications targeting narrow problems; calibrated support preventing over-reliance; preserved human judgment maintaining expertise; and enterprise-wide prompt engineering training. Wilson and Daugherty (2018) propose humans training, explaining, and responsibly deploying AI while AI augments cognition and performs tasks.

The literature identifies predominant augmentative patterns where AI supports iterative refinement and knowledge acquisition. For complex tasks, this implies designing co-creation cycles, explanatory dialogues, and verification mechanisms positioning AI as collaborative partner (Accenture 2025, Handa, Tamkin et al. 2025). "Copiloting" exemplifies this complexity, structuring work around natural language interactions leveraging AI's content creation capacity (Banh, Holldack et al. 2025).

Trust emerges as collaboration's cornerstone. The "black box" nature of sophisticated models erodes confidence (Floridi 2023). Explainable AI addresses this through human-centred explanations, proving successful in high-stakes fields (Bilal, Ebert et al. 2025). Building trust extends beyond explainability to privacy, bias, and control concerns. Trustworthy AI frameworks emphasize human oversight, demonstrated competence, and uncertainty reduction (Li, Wu et al. 2024). Multi-stage reasoning techniques like Introspective Uncertainty Quantification enable AI self-critique, providing reliable uncertainty estimates (Mei, Zhang et al. 2025).

Despite these advances, significant risks persist. AI may facilitate labour exploitation through "growth without calories" where firms expand without hiring (Galloway 2024). Over-reliance risks cognitive degradation, undermining motivation and critical thinking (Singh, Guan et al. 2025, Wu, Liu et al. 2025). The human role must be reframed as "steward" actively verifying outputs and refining prompts (Collins, Bhatt et al. 2025, Lee, Sarkar et al. 2025).

## 3. Methodology

### 3.1 Research Philosophy and Design

This study employed a pragmatic mixed-methods design to comprehensively investigate AI's impact on Strategic Technology Management. The methodology was anchored by critical realist ontology, acknowledging external reality independent of perception while recognizing that human understanding shapes knowledge of that reality (Saunders and Lewis 2017). This philosophical stance enabled exploration of both objective organizational structures and subjective individual experiences within those structures.

The research adopted a sequential explanatory mixed-methods approach, beginning with quantitative data collection to establish broad foundations, followed by qualitative inquiry providing contextual depth (Creswell 2021). This structure ensured initial quantitative findings informed and directed subsequent qualitative investigation, yielding richer, more nuanced understanding. The design's robustness stemmed from its ability to achieve explanatory power unattainable through either method in isolation.

### 3.2 Quantitative Phase: Online Survey

The initial quantitative phase involved an online survey administered via Google Forms between November 2024 and February 2025. The instrument was designed to verify propositions and gaps identified in existing literature while establishing the prevalence of key phenomena across the broader professional population.

*3.2.1 Survey Design*

The survey comprised 21 sections containing 120 single-choice questions, predominantly utilizing 5-point Likert scales. Sections addressed themes including AI adoption in STM practices, trust levels, cultural readiness, human-AI interaction, resources, skills, and investment priorities. Questions were structured to operationalize abstract constructs such as perceived workload associated with AI output verification, trust in AI systems, and organizational AI governance preparedness. The instrument underwent pilot testing to ensure question neutrality and minimize response bias.

### 3.2.2 Sampling and Data Collection

The study employed convenience and snowball sampling through professional networks. Initial distribution to LinkedIn Groups reached potential audience of 7 million accounts, yielding limited responses. Subsequently, distribution to the researcher's personal LinkedIn network (1,726 users) proved more effective, generating 278 responses with 230 valid completions (82.7% validity rate).

Participants were predominantly Europe-based (81%), with professional backgrounds concentrated in Information Technology, Software Development, and consultancy (70% collectively). Respondent roles split between consultancy (63%) and leadership positions (37%), providing dual perspectives from technology strategy advisors and decision-makers.

### 3.2.3 Statistical Analysis

Quantitative data underwent rigorous statistical analysis including:

- Reliability testing using Cronbach's Alpha ($\alpha = 0.906$ for 53 items), demonstrating high internal consistency
- Descriptive statistics calculating means and standard deviations for Likert-scale items
- Pearson correlation analysis investigating relationships between variables
- Linear regression analysis examining predictive relationships

Statistical analysis revealed significant relationships, including positive correlation between AI's strategic action definition role and uncertainty handling ability ($r=0.52$, $p<0.01$), and predictive relationship between perceived AI effectiveness and future adoption intentions ($\beta=0.63$, $p<0.001$).

## 3.3 Qualitative Phase: Expert Interviews

Following quantitative analysis, the study transitioned to qualitative inquiry using semi-structured interviews with 14 industry experts. This phase employed purposive sampling to select participants providing rich, contextual insights based on specific expertise.

### 3.3.1 Participant Selection

Four distinct professional groups were targeted:

1. **Chief Technology Officers (CTO) / Chief Information Officers (CIO) / IT Directors (n=9):** Senior technology leaders making strategic decisions about technology adoption and resource allocation
2. **Academic Researcher (n=1):** Professor of Technology and Innovation Management specializing in technology strategy roadmap digitalization
3. **Strategic Portfolio Management Product/Service Vendors (n=3):** Professionals from companies developing Strategic Portfolio Management tools embedding AI
4. **AI Platform Architect (n=1):** Architect from leading Generative AI platform company

Response rates varied across groups, with approximately 10% of contacted CTOs/CIOs participating, reflecting expected executive-level engagement rates while yielding sufficient data for rich qualitative analysis.

### 3.3.2 Interview Protocol

The semi-structured interview guide was crafted based on survey findings, enabling focused exploration of emergent themes. The protocol addressed six key themes:

- AI adoption and readiness in strategic management
- Critical resources for AI integration in technology strategy decision-making
- Speculative views on autonomous AI in STM (the "Agentic CTO" concept)
- Balance between human expertise and AI capabilities
- Strategic roadmap formulation under VUCA conditions
- Trust, accountability, and ethical concerns

Interviews were conducted remotely via Microsoft Teams between April and June 2025, with sessions digitally recorded and transcribed using automated services. The semi-structured format allowed flexibility to explore emergent topics while maintaining consistency across core themes.

## 3.4 Data Analysis and Integration

### 3.4.1 Qualitative Analysis

Interview transcripts underwent systematic analysis using NVivo software:

- Initial open coding generated 651 distinct codes
- Codes were clustered according to research questions
- Further refinement aligned codes with AIbSTM framework categories
- Final analysis produced 20 refined categories directly linked to research questions

To ensure coding reliability, an independent researcher reviewed 20% of coded data, achieving 70% inter-coder agreement (kappa coefficient), indicating strong consensus and validating analytical consistency.

*3.4.2 Data Triangulation*

Integration of quantitative and qualitative data occurred at the interpretation stage. The mixed-methods design enabled meta-inferences (Tashakkori and Teddlie 2010) greater than individual parts' sum. Quantitative findings provided broad landscape understanding, while qualitative data offered contextual depth necessary for interpretation. For example, survey results showing low AI adoption for strategic functions were explained through interviews revealing organizational barriers and cultural resistance.

This triangulation enhanced validity and reliability by ensuring patterns observed in one dataset could be corroborated or contextualized by another. The convergence of findings from diverse sources provided robust, well-substantiated conclusions about AI's impact on STM.

## 3.5 Ethical Considerations

The research adhered to strict ethical protocols ensuring participant rights and data privacy:

- **Informed Consent:** All participants received clear explanations of research objectives, data use, and their rights
- **Confidentiality:** Personal identifiers were removed during analysis; pseudonyms protected interview participants' identities
- **Data Security:** All materials stored on password-protected, encrypted drives with restricted access
- **Transparency:** Research aims and methods clearly communicated to all participants

Ethical framework reflected the study's emphasis on human-centred approaches prioritizing thoughtful AI tool integration within frameworks emphasizing ethical oversight.

## 3.6 Limitations

While providing comprehensive exploration, the study acknowledges methodological limitations:

- Convenience and snowball sampling may introduce selection bias
- Reliance on self-reported survey data presents response bias risk

- LinkedIn-based recruitment potentially overrepresents technology-forward professionals
- Small interview sample size limits generalizability of qualitative findings
- Cross-sectional design captures single time point rather than longitudinal evolution

These limitations were addressed through mixed-methods triangulation and transparent acknowledgment in findings interpretation.

## 4. Findings and Analysis

### 4.1 Strategic Alignment and AI Innovation

#### *4.1.1 Survey Findings on AI's Role in STM*

The quantitative analysis revealed AI's moderate yet evolving impact on strategic technology roadmap formulation. AI demonstrated strongest value in providing leadership advisory functions (M=3.24, SD=0.99), suggesting effectiveness in integrating data-driven insights into strategic formulation. However, AI's capability to navigate VUCA scenarios remained limited (M=3.08, SD=1.04), indicating current inability to fully manage unpredictability inherent in uncertain environments.

AI proved most effective in tactical roadmap aspects, particularly defining "Know WHAT" (strategic actions) (M=3.47, SD=1.00) and "Know HOW" (implementation pathways) (M=3.54, SD=1.00). Conversely, effectiveness diminished for higher-level strategic elements including "Know WHY" (strategic intent) (M=3.14, SD=1.00) and "Know WHEN" (timing) (M=3.00, SD=1.00), highlighting limitations in contextual reasoning and strategic timing.

A significant barrier emerged in data availability and quality (M=4.06, SD=1.00), directly impacting AI's capacity for reliable recommendations under uncertainty. Strong positive correlation existed between AI's strategic action definition role and uncertainty handling ability (r=0.52, p<0.01), with highest correlation observed for "Know WHY" (r=0.49, p<0.01), emphasizing effectiveness in identifying strategic goals as key factor in leveraging AI for informed decision-making.

Organizational challenges significantly impeded adoption, including limited leadership mandates (M=2.75, SD=1.00) and organizational resistance to change (M=3.48, SD=1.00). Despite current limitations, respondents expressed optimism about AI's future potential for VUCA management within 1-2 years (M=3.84, SD=0.85).

#### *4.1.2 Expert Perspectives on Strategic Alignment*

Interview analysis revealed AI's role transcends simple optimization, functioning as catalyst for achieving unprecedented strategic clarity. Experts consistently emphasized the imperative of defining clear business purposes before AI deployment. As one CTO stated: "Creating what I call a business architectural definition of what it is that I'm trying to achieve" is fundamental to successful alignment.

AI necessitates shifting investment logic from traditional ROI metrics toward enabling previously unachievable strategic outcomes. An expert noted AI enables companies to "achieve goals that were not achievable before," representing a "huge opportunity because there are things that could not be done before." This shift demands rigorous linkage between investments and defined business capabilities, avoiding scattered efforts pursuing "silver bullet" solutions without maturation.

However, interviews also revealed AI's profound limitations in authentic strategic cognition. Multiple experts warned of AI's inability to capture subjective strategic elements - the "air" in strategic workshops encompassing unspoken nuances, cultural DNA, and accumulated leadership intuition. One academic cautioned that AI-generated strategies risk being "inauthentic and disingenuous," technically correct but lacking embedded human context vital for true alignment.

The "Agentic CTO" concept, inspired by NVIDIA CEO Jensen Huang's vision of AI agents replacing human technology leadership (NVIDIA 2025), was largely rejected by interviewees. One expert suggested this narrative primarily serves commercial agendas: "I think it's useful for him to maintain a certain narrative that drives consumption of GPUs." Another distinguished between operational automation feasibility and strategic planning complexity: "There's an awful lot of standard automation that can be done… but to get to a point where you fit into process understanding… that's not feasible."

## 4.2 Resources and Capabilities Through the RBV Lens

### 4.2.1 Quantitative Assessment of AI Resources

Survey findings revealed clear prioritization hierarchies among AI resources. Computational power emerged as commodity resource (M=1.62, SD=1.16), with low strategic value indicating "Buy" orientation through cloud providers. Conversely, three core resources emerged as significant differentiators requiring "Build" orientation:

- **Proprietary Data** (M=3.81, SD=1.33): Highest valued resource due to uniqueness and context-specificity
- **Human Talent** (M=3.68, SD=1.27): Valued for specialized, tacit knowledge translating generic AI tools into strategic assets
- **AI Governance** (M=3.83, SD=1.36): Complex, organization-specific capability building trust and ensuring responsible deployment

Despite high perceived value, current governance investments showed concerning gaps (M=3.20, SD=1.06), though future investment intentions were strong (M=3.88, SD=0.88).

Algorithms occupied middle ground (M=3.07, SD=1.50), suggesting strategic value contingent on integration with proprietary resources.

Investment priorities for next 1-2 years clearly focused on Data (M=3.75, SD=1.29) and Human Talent (M=3.54, SD=1.35), while Computational Power (M=2.54, SD=1.41) and surprisingly AI Governance (M=2.89, SD=1.42) received lower priority despite recognized importance.

### 4.2.2 Qualitative Insights on Critical Capabilities

Expert interviews validated and expanded survey findings regarding essential resources and capabilities. Data accessibility emerged as foundational, with one CIO emphasizing: "It is data accessibility that is critical. We have structured governance to direct AI initiatives and ensuring data is accessible and of high quality."

Interviews revealed critical tension between open knowledge sharing needs and proprietary data protection imperatives. One expert proposed radical accessibility: "To achieve AI's Nirvana… We would have to create a world where we're accepting that our LLM knowledge can be publicly shared with anyone," while others emphasized protective approaches for maintaining competitive advantage.

Governance emerged as particularly complex capability. Interviews highlighted leadership disengagement as major barrier, with one expert noting "failure at the top" in understanding and embracing technology governance responsibilities. Effective governance requires dynamic capabilities including integrated feedback loops, adaptive compliance frameworks balancing ethical standards across jurisdictions, and continuous oversight mechanisms.

Human talent discussions emphasized "fusion skills" combining domain expertise with AI literacy. Middle management emerged as critical bridge, requiring AI skills to facilitate communication between adopters and inhibitors. Organizations need comprehensive approaches including targeted hiring, strategic reskilling, and fundamental workforce reconfiguration.

Technical orchestration capabilities proved essential beyond basic resource access. This includes ability to unify structured/unstructured data at scale, transform core processes for strategic advantage, and manage current AI limitations including hallucinations, unpredictable costs, and rapidly evolving models. One expert highlighted scalability concern: "Can the technology support the volume? How would we be able to scale up?"

## 4.3 Human-AI Interaction Design

### *4.3.1 Survey Evidence on Collaboration Patterns*

Quantitative findings revealed cautiously optimistic yet constrained views of current human-AI collaboration. Majority acknowledged AI's enhancement potential, particularly through "only when needed" support (M=3.81, SD=0.80), framing AI as valuable augmentation tool. However, path to deeper symbiotic relationships faced significant barriers including ethical concerns (M=3.30, SD=1.04) and interaction difficulties, tempering expectations for rapid evolution toward true symbiosis within 1-2 years (M=2.99, SD=1.09).

Practitioners demonstrated proactive skill development stance (M=3.99, SD=0.78) with moderate confidence interpreting AI outputs (M=3.48, SD=0.92), though progress was impeded by resource constraints (M=3.43, SD=0.99). Strong strategic commitment emerged for future investments in human-AI capabilities (M=3.99, SD=0.89), with significant predictive relationship between perceived AI effectiveness and future adoption ($\beta=0.63$, $p<0.001$).

Trust-building efforts through transparent policies (M=3.47, SD=1.05) and ethical frameworks (M=3.61, SD=1.00) showed positive correlation with adoption ($r=0.43$, $p<0.01$). However, critical trust deficits persisted due to audit implementation challenges (M=3.40, SD=0.89). Practitioners overwhelmingly rejected accountability delegation to AI (M=2.13, SD=1.01), citing inadequate legal frameworks (M=3.37, SD=1.05), reinforcing human oversight requirements.

### *4.3.2 Expert Views on Augmentation Models*

Interview analysis consistently positioned AI as strategic augmenter rather than replacement. Experts described AI's value in automating labour-intensive groundwork, with one noting AI delivers "75-90% of drafting" while requiring human oversight for contextualization within organizational nuances and strategic intent.

Iterative co-creation emerged as critical theme. AI serves as dynamic workshop participant providing real-time insights and challenging assumptions while humans refine direction and make final judgments. This symbiosis ensures AI's analytical speed complements human intuition, particularly for ambiguous decisions.

Clear boundaries delineating AI's supportive role proved essential. While AI aids research and optimization, even acting as "customer viewpoint" proxy, it cannot lead teams or assume accountability. Human leadership remains essential for interpreting insights, navigating organizational politics, and driving culturally aligned execution.

Trust emerged as paramount concern, with "black box" AI nature creating fundamental barriers. Building trust requires transparency, consistent performance, and user involvement in collaborative design. As one expert emphasized need for "celebrating successes" to build familiarity and confidence. The necessity of explainable AI and verification mechanisms was consistently stressed, with experts advocating for low-risk pilots and co-creation approaches overcoming fears.

Cultural transformation requirements featured prominently. Organizations must cultivate critical thinking skills for evaluating AI outputs and formulating precise questions. One expert stressed importance of "being able to articulate those questions" to ensure contextually appropriate AI responses. Value realization requires shifting from hype to measurable outcomes, strategically offloading routine tasks to free human talent for higher-order strategic synthesis.

## 5. Discussion of Findings

### 5.1 RQ1: AI Innovation in STM Success Factors

The findings establish that AI fundamentally innovates two critical success factors for STM roadmap formulation under uncertainty and creativity. First, AI enables a paradigm shift from intuition-based to evidence-based strategic alignment, providing data-driven insights that compel organizations to define clear business purposes and robust architectures before implementation. This finding challenges the survey's surface-level indication that AI struggles with higher-level strategic elements, instead revealing that AI acts as a catalyst forcing strategic clarity.

The transformation in strategic alignment represents more than technical optimization. As evidenced by both quantitative and qualitative data, AI's ability to process vast datasets and generate predictive insights addresses the long-standing challenge of cognitive bias in strategic planning. The correlation between AI's effectiveness in defining strategic actions and handling uncertainty ($r=0.52$) demonstrates its potential for managing volatile environments. However, the expert interviews provide crucial context - this potential is realized only when organizations embrace the discipline AI demands.

Second, AI facilitates continuous adaptation by transforming the static STM roadmap into a dynamic, living guide. The survey's optimism about future VUCA management capabilities ($M=3.84$) aligns with experts' descriptions of shortened planning horizons and iterative strategy cycles. This shift from episodic to continuous strategic planning represents a fundamental reconceptualization of STM, where the "infinite game" approach replaces rigid multi-year plans with fluid, adaptive frameworks.

The significance of these innovations extends beyond process improvements. By forcing purpose-driven clarity and enabling continuous adaptation, AI addresses the core problem identified in the introduction - the persistent failure of technology investments to deliver expected value. The shift from ROI-focused metrics to purpose-driven investment logic, validated across both data sources, suggests AI's role in ensuring technology investments align with genuine strategic outcomes rather than following trends or pursuing isolated efficiencies.

However, the research also reveals significant limitations. The "mirage of alignment" warning from experts - where AI generates superficially impressive but ultimately inauthentic strategies - highlights the persistent importance of human judgment. The gap between AI's tactical strengths (Know WHAT/HOW) and strategic weaknesses (Know WHY/WHEN) cannot be fully bridged by technology alone, requiring human strategists to provide contextual intelligence and navigate organizational complexities AI cannot grasp.

## 5.2 RQ2: Essential Resources and Capabilities

The research provides a comprehensive answer to what resources and capabilities organizations require for AI-enhanced STM, revealing a clear hierarchy that challenges conventional assumptions about AI adoption. The findings definitively establish that competitive advantage stems not from technological resources themselves but from organization-specific capabilities to orchestrate them effectively.

The commoditization of computational power, despite its necessity, represents a critical insight with profound strategic implications. Organizations pursuing competitive advantage through computational accumulation are fundamentally misallocating resources. Instead, the research identifies three categories of truly differentiating resources: proprietary data ecosystems, specialized human talent, and robust governance capabilities. Each meet VRIN criteria in distinct ways that commodity compute cannot achieve.

The underinvestment in governance despite its recognized importance reveals a significant implementation gap. This paradox - where organizations acknowledge governance as differentiating capability yet fail to prioritize investment - suggests deeper organizational challenges. Expert interviews illuminate these barriers: leadership disengagement, rapidly evolving technology outpacing governance frameworks, and the inherent difficulty of governing poorly understood systems. The finding that governance may transition from differentiator to market prerequisite (exemplified by the EU AI Act) adds urgency to addressing this gap.

Human talent emerges as more nuanced than simple technical expertise. The concept of "fusion skills" - combining domain knowledge with AI literacy - reframes workforce development from hiring data scientists to cultivating hybrid competencies across the organization. Middle management's critical bridging role between AI adopters and sceptics represents an overlooked capability that determines implementation success.

The surprising undervaluation of algorithms in both survey and interview data, contrasting with literature emphasis on pre-trained domain models, suggests a potential blind spot in organizational perception. This finding implies organizations may be overlooking strategic opportunities in developing or acquiring specialized algorithmic capabilities tailored to their specific contexts.

## 5.3 RQ3: Designing Human-AI Interaction

The research provides a multifaceted answer to how human-AI interaction should be designed for complex STM tasks, definitively establishing that effective design requires deliberate, continuous processes prioritizing augmentation over automation. The convergence of survey and interview findings on this point is striking - despite technological possibilities for greater automation, the optimal design maintains human centrality.

The augmentation paradigm emerging from the data positions AI as a "cognitive partner" handling foundational analysis while preserving human responsibility for judgment, creativity, and accountability. The consistent finding across both data sources that AI can automate 75-90% of groundwork while requiring human oversight for the remaining critical percentage establishes a clear division of labour. This is not merely a technical arrangement but a fundamental reconceptualization of strategic work.

Trust emerges as the binding constraint on effective human-AI collaboration. The triangulation of survey data showing trust deficits despite policy implementation with interview insights on "black box" anxieties reveals that trust cannot be mandated through governance alone. Instead, it must be cultivated through transparency, consistent performance, and experiential learning. The emphasis on "celebrating successes" and low-risk pilots suggests trust-building as an iterative, social process rather than a technical challenge.

The overwhelming rejection of AI accountability (M=2.13) coupled with inadequate legal frameworks reinforces that human-AI interaction design must preserve clear lines of human responsibility. This finding definitively answers speculation about autonomous AI leadership - the "Agentic CTO" concept is neither technically feasible nor organizationally desirable given current capabilities and frameworks.

The cultural transformation requirements identified across both data sources highlight that effective human-AI interaction extends beyond interface design to fundamental organizational change. The need to cultivate critical thinking, precise questioning, and value-focused integration suggests that organizations must develop new cultural competencies alongside technical capabilities. The shift from "doing the same" to "doing more" with AI represents not just operational improvement but strategic transformation of human work itself.

# AIbSTM FRAMEWORK current and speculative View

## RESEARCH QUESTIONS

**RQ#1:** What critical success factors does AI innovate for formulating the strategic technology management roadmap under conditions of creativity and uncertainty?

**RQ#2:** What resources and capabilities should organisations have to enhance their strategic technology management roadmap formulation in complex scenarios?

**RQ#3:** How should the interaction between humans and AI be designed to enhance the performance of complex strategic technology management tasks?

## AI BASED STM FRAMEWORK

**CURRENT**

| Strategic Alignment | Continuous Adaptation | Resource-Based View | Leadership and Culture | AI – Human Interaction |
|---|---|---|---|---|
| • Data-Driven Advisory<br>• Action Definition (Know WHAT)<br>• Pathway Determination (Know HOW) | • Operational Workflow Integration<br>• Incremental Confidence Building<br>• VUCA Preparedness Catalyst | • Data Differentiation<br>• Governance Advantage<br>• Talent Orchestration | • Cultural Diagnostic Tool<br>• Operational Task Automation | • Augmentation Effectiveness<br>• Skill Development Engagement<br>• Leadership-Driven Collaboration |

**SPECULATIVE**

| Strategic Alignment | Continuous Adaptation | Resource-Based View | Leadership and Culture | AI – Human Interaction |
|---|---|---|---|---|
| • Agentic CTO Realization<br>• Human Oversight Evolution<br>• AI Forces Strategic Clarity | • Infinite Strategy Refinement<br>• Human-AI Co-Evolution<br>• Fluid Operating Models | • Radical Data Sharing<br>• Adaptive Governance Systems<br>• Risk Capitalization Capability | • Cognitive Offloading Assistants<br>• Agentic Marketplace Integration<br>• Value-Driven Organizational Redesign | • Symbiotic Workflow Integration<br>• Organizational Restructuring<br>• Engineered Trust via Co-Creation |

*Figure 1 - AIbSTM Framework highlighting the current practices across each dimension of the framework and the speculative view for evolutionary directions*

## 5.4 Synthesis and Implications

The integrated findings across all three research questions reveal a coherent narrative about AI's role in STM. Rather than replacing human judgment or automating strategy, AI emerges as a powerful catalyst that simultaneously demands and enables higher-order human contributions. Technology forces organizational clarity while providing tools for continuous adaptation, requires new resources and capabilities while commoditizing others, and augments human capacity while preserving human accountability.

The AIbSTM framework, visualized in Figure 1, synthesises these insights by providing both theoretical advancement and practical guidance. Its three-layered structure - strategic alignment/continuous adaptation, resource-based view, and leadership/culture/interaction - maps directly onto the research questions while offering actionable pathways for implementation. The framework's value lies not just in describing current practices but in prescribing the organizational transformations necessary for successful AI integration.

The persistent gap between potential and realization identified across all dimensions points to a fundamental challenge: the cognitive and socio-technical chasms preventing full AI utilization. These chasms are not merely implementation barriers but represent fundamental disconnects between AI's data-driven logic and human contextual intelligence, between technological capabilities and organizational readiness, between automation potential and accountability requirements.

## 6. Conclusions and Implications

### 6.1 Theoretical Contributions

This research makes three significant theoretical contributions to strategic management and organizational theory. First, it extends the Resource-Based View to AI contexts by reconceptualizing competitive advantage sources. Rather than viewing AI as a resource to be acquired, the study establishes that advantage derives from organization-specific capabilities to integrate, manage, and evolve AI systems. This includes developing ethical governance frameworks, cultivating dynamic leadership, and achieving deliberate integration within human-centred frameworks. The finding that computational power is commoditized while data, talent, and governance remain differentiating fundamentally reframes how RBV applies to digital resources.

Second, the research addresses the socio-technical chasm in AI adoption through detailed examination of trust, cultural readiness, and governance requirements. By identifying specific mechanisms through which organizations can bridge gaps between technical potential and human reality, the study provides a theoretical framework for understanding why technically superior AI solutions often fail to deliver organizational value. The emphasis on trust as an iteratively constructed social phenomenon rather than a technical feature advances understanding of technology adoption in complex organizational contexts.

Third, the identification and analysis of the cognitive chasm between human strategic thinking and AI's data-driven processing provides a novel lens for understanding human-AI collaboration dynamics. The framework encouraging symbiotic relationships where AI provides predictive power while humans provide contextual nuance addresses the fundamental challenge of algorithmic over-reliance while preserving essential human contributions to strategy formulation.

## 6.2 Practical Implications

The AIbSTM framework offers practitioners a comprehensive blueprint for navigating AI integration in strategic technology management. Its three-layered structure provides specific, actionable guidance:

For strategic alignment, organizations should use AI to transition from static planning to continuous adaptation, employing real-time analytics and scenario modelling to maintain dynamic alignment with business objectives. This requires shifting investment logic from traditional ROI metrics toward enabling previously unattainable strategic outcomes.

Regarding resources and capabilities, the framework prescribes prioritizing investments in data governance and curation, developing fusion skills combining domain expertise with AI literacy, fostering cross-functional collaboration, and treating AI limitation management as a core capability. The clear identification of proprietary data, specialized talent, and governance as differentiating resources while computational power remains commoditized provides clear investment priorities.

For human-AI interaction, the framework emphasizes designing augmentation workflows where AI handles foundational analysis while humans retain responsibility for judgment and creativity. Organizations must cultivate critical thinking skills, establish clear accountability frameworks, promote bidirectional learning, and develop data-centric, agile cultures supporting continuous adaptation.

The framework's practical value extends beyond prescriptive guidance to enabling organizational self-assessment. Using the AIbSTM conceptual framework checklist defined in Table 1, leaders can evaluate readiness across multiple dimensions, identify capability gaps, and prioritize interventions ensuring AI investments yield strategic rather than merely tactical value.

*Table 1 - Checklist and speculating view for the AIbSTM conceptual framework dimensions for AI facilitating STM*

| Strategic Alignment | | |
|---|---|---|
| Observed behaviours and checklist | **Data-Driven Leadership Advisory** - AI provides insights to leadership, informing strategy decisions through analysed data<br><br>Validated by **AI-powered drafted strategy**: using generative AI to streamline initial strategic planning and documentation. This includes crafting tailored strategies, goals, and objectives based on user input, which allows for more adaptable workflows and cross-verification of information. | |
| | **Action Definition (Know WHAT)** - AI identifies specific strategic actions needed based on analysis<br><br>Validated by **Scenario simulation**: simulate scenarios like market shifts and competitor actions to make strategies more adaptable and proactively test strategic options. | |
| | **Pathway Determination (Know HOW)** - AI outlines implementation methods and technological pathways for strategic initiatives<br><br>Validated by **Automated portfolio management**: enablement of [or deployment of tools to provide] AI based support portfolio management by providing real-time insights, modelling what-if scenarios, and optimizing portfolios for strategic prioritization. | |
| Speculating – how AI could support higher strategic alignment | **Agentic CTO Realization** - AI autonomously performing core CTO functions (strategy synthesis, roadmap creation). More feasible for specific hierarchical/digital-native cultures with low ethical constraints | |
| | **Human Oversight Evolution** - Human roles shift from direct control to phased supervision of AI agents. Requires structured roadmaps (from controls to supervision) and tolerates non-linear progression | |
| | **AI Forces Strategic Clarity** - AI reinforcing unprecedented precision in business purpose, objectives, and translation to technology. Still success depends on robust organization' business architecture and overcoming measurement | |
| **Continuous Adaptation** | | |
| Observed behaviours and checklist | **Operational Workflow Integration** - AI effectively handles routine tasks, establishing a foothold in daily processes despite strategic hesitancy<br><br>Validated by Integrate **AI into Workforce Planning**: using AI for scenario planning to anticipate skill gaps, automate talent allocation, and identify reskilling pathways. | |
| | **Incremental Confidence Building** - Gradual AI adoption in low-risk areas fosters organizational trust for future strategic scaling<br><br>Validated by Integrate **Human-AI Bidirectional Learning**: creating a continuous feedback loop where AI adapts to human feedback, and humans adjust to AI insights. | |
| | **VUCA Preparedness Catalyst** - AI enables early-stage volatility monitoring, providing data-driven signals for proactive adaptation.<br><br>Validated by Integrate **Dynamic What-if Scenarios**: using AI to model and test strategic options, ensuring continuous adaptation to market and competitor changes. | |
| Speculating – how AI could define and facilitate continuous adaptation | **Infinite Strategy Refinement** - AI could enable real-time roadmap updates weekly/monthly. Requires human oversight to prevent "superficial outputs" | |
| | **Human-AI Co-Evolution** - AI as symbiotic partner navigating VUCA uncertainties. "Cultural inertia" may obstruct equitable collaboration | |
| | **Fluid Operating Models** - Self-adjusting workflows where AI reconfigures strategies dynamically. Demands "structured change management" to avoid tool redundancy. | |
| **Resource-Based View** | | |
| Observed behaviours and checklist | **Data Differentiation** - Proprietary data is leveraged as a rare, valuable asset for unique AI insights, enabling competitive differentiation.<br><br>Validated by **Integrate AI Resources and Capabilities Checklists**: to evaluate the key resources and capabilities as proprietary data ecosystems, specialized human talent, and robust governance capabilities. | |
| | **Governance Advantage** - Robust AI governance is a non-substitutable capability building trust, mitigating unique risks, and enabling responsible, strategic AI deployment.<br><br>Validated by **AI Governance model**: to manage data workflows and establish ethical frameworks, roles, and compliance. | |
| | **Talent Orchestration** - Specialized human talent possesses tacit knowledge crucial for contextually developing, deploying, and managing AI effectively within the organization.<br><br>Validated by **Hybrid Talent model**: integrate "fusion" skills with AI capabilities for strategic alignment. | |

| | |
|---|---|
| Speculating – what new resources will determine a competitive advantage in an AI driven STM | **Radical Data Sharing** - Public sharing of proprietary data/knowledge (e.g., "publicly shared LLM knowledge") could unlock unprecedented AI value, redefining exclusivity. This requires balancing conservative views who fiercely protect proprietary data ("very protective") as a core strategic asset. |
| | **Adaptive Governance Systems** - Dynamic, embedded feedback loops and "polished frameworks" will continuously evolve governance, replacing static compliance. Requires overcoming leadership disengagement ("failure at the top") and ethical ambiguity for effective implementation. |
| | **Risk Capitalization Capability** - Organizations will strategically cultivate "healthy risk appetite" as a core capability, tolerating operational risk for speed/innovation. Must rigorously balance this with mitigating existential threats (e.g., breaches, reputational damage) highlighted as critical. |
| **Leadership and Culture** | |
| Observed behaviours and checklist | **Cultural Diagnostic Tool** - AI monitors organizational culture, identifies alignment/friction points via sentiment analysis<br><br>Validated by **Culture and Leadership Diagnostic Tool**: assess and guide organizational culture and leadership to effectively manage AI integration and human-AI interactions. |
| | **Operational Task Automation** - AI handles routine operational "dirty work", freeing human capacity for higher-value activities<br><br>Validated by **Value Creation guidelines**: define the value tree, linking business objectives to measurable key performance indicators (KPIs). |
| Speculating – how AI could change strategic technology leadership and culture | **Cognitive Offloading Assistants** - AI as "digital mentors/coaches" elevating human work. Requires workflow redesign to prevent value loss |
| | **Agentic Marketplace Integration** - Marketplaces enable accessible AI agents for task execution. It requires managing "Extremely messy" unvetted agents risk ethics/alignment. |
| | **Value-Driven Organizational Redesign** - AI enables human-AI workflows focusing talent on strategy. It requires managing resistance to structural change hinders adoption |
| **AI – Human Interaction** | |
| Observed behaviours and checklist | **Augmentation Effectiveness** - AI successfully automates groundwork (data/code generation), freeing humans for strategy/validation |
| | Validated by **Design for co-creation**: design for co-creation cycles, explanatory dialogues, and verification mechanisms, positioning AI as a collaborative partner. |
| | **Skill Development Engagement** - High practitioner upskilling activity driven by leadership mandates on critical AI competencies.<br><br>Validated by **Establish human judgment checkpoints**: preserve human expertise by creating verification points for AI-generated outputs. |
| | **Leadership-Driven Collaboration** - Leaders actively promote AI-human collaboration to boost decision-making performance in specific workflows.<br><br>Validated by **Checklist for AI-Assisted Decision Making**: integrate AI to mitigate human bias, enhance foresight, and improve strategic decision-making. It should include validation for Bias Mitigation; Enhanced Foresight; Data and Insights; Interactive Loop; Dynamic Roadmaps. |
| Speculating – how AI and human will engage in an AI driven STM | **Symbiotic Workflow Integration** - AI evolves into a dynamic co-creator, providing real-time insights and challenging assumptions iteratively within tasks. It requires removing ethical and trust barriers. |
| | **Organizational Restructuring** - Shift to "diamond-shaped" teams with AI agents handling execution, reducing low-level roles and enabling fractional leadership. It requires moving [partial] accountability from human to machine developing new legal frameworks covering AI responsibility. |
| | **Engineered Trust via Co-Creation** - Trust built through explainable AI, low-risk pilots, user co-design, and celebrating joint successes to overcome "black box" fears. It requires reducing effectiveness gaps; and to address ethics concerns reducing trust. |

## 6.3 Limitations

This study acknowledges several limitations affecting its generalizability and applicability. The convenience and snowball sampling approach may introduce selection bias, with findings potentially overrepresenting technology-forward professionals active on LinkedIn. The cross-sectional design captures a single temporal snapshot of a rapidly evolving phenomenon, limiting insights into evolutionary dynamics of AI adoption.

The small sample size for qualitative interviews (n=14), while providing rich insights, constrains generalizability of expert perspectives. Low response rates from certain groups, particularly AI platform professionals and academics, may have resulted in underrepresentation of important viewpoints. Additionally, the study's grounding in current technological capabilities means findings may require reassessment as AI capabilities advance.

The absence of longitudinal data prevents examination of how trust, capabilities, and organizational structures evolve over time with AI integration. The research also lacks experimental validation of proposed frameworks, relying instead on reported experiences and perceptions which may be subject to various biases.

## 6.4 Future Research Directions

Several promising avenues for future research emerge from this study. Empirical validation of the AIbSTM framework through comparative case studies across industries would strengthen its generalizability and identify context-specific adaptations. Longitudinal research tracking trust evolution and capability development would provide insights into the dynamics of human-AI collaboration maturation.

Experimental studies, including lab-based "Agentic CTO" prototypes and simulated workshops examining AI's impact on group decision-making, could validate theoretical propositions under controlled conditions. Research into the organizational implications of AI adoption, including structural reconfigurations and new governance models, would address the implementation challenges identified in this study.

The undervaluation of algorithmic resources despite their theoretical importance warrants dedicated investigation into how organizations can develop or acquire specialized algorithmic capabilities. Studies examining the transition of governance from differentiating capability to market prerequisite would illuminate how regulatory frameworks shape competitive dynamics.

Investigation of cultural and geographic variations in AI adoption patterns would enhance understanding of how different contexts influence human-AI collaboration models. Finally, research into mitigation strategies for cognitive and socio-technical chasms would provide practical pathways for organizations struggling to realize AI's full potential.

## 6.5 Conclusion

This research establishes a comprehensive framework for understanding AI's transformative impact on strategic technology management. By addressing the persistent challenge of underperforming technology investments, the study demonstrates that AI's value lies not in automating strategy but in catalysing fundamental organizational transformation. The findings definitively show that successful AI integration requires moving beyond technical implementation to address cognitive and socio-technical challenges through deliberate cultivation of human-AI partnerships.

The rejection of autonomous AI leadership concepts in favour of human-centric augmentation represents a crucial finding with immediate practical implications. Organizations pursuing competitive advantage through AI must focus not on replacing human judgment but on creating synergistic relationships where AI's analytical power complements human creativity, contextual intelligence, and ethical reasoning.

The AIbSTM framework provides both theoretical advancement and practical guidance for navigating this transformation. By identifying specific resources, capabilities, and interaction designs necessary for success, the research offers a clear pathway for organizations seeking to enhance their strategic technology management through AI while avoiding common pitfalls of over-automation or under-utilization.

Ultimately, this research matters because it provides evidence-based guidance for a significant organizational challenge: integrating artificial intelligence into strategic decision-making in ways that enhance rather than diminish human contribution. As organizations increasingly rely on AI for competitive advantage, understanding how to design effective human-AI partnerships becomes not merely important but essential for survival and success in the digital age.

The journey toward AI-enhanced strategic technology management is not one of linear technological progress but of continuous co-evolution between human and artificial intelligence. This research provides the theoretical foundation and practical tools for organizations to navigate this journey successfully, ensuring technology investments deliver genuine strategic value while preserving the essentially human elements of creativity, judgment, and accountability that remain irreplaceable in strategic leadership.